\documentclass[%
 reprint,
 amsmath,amssymb,
 aps,
]{revtex4-2}

\usepackage{graphicx}
\usepackage{dcolumn}
\usepackage{bm}
\usepackage{xcolor}

\begin{document}

\preprint{APS/123-QED}

\title{General Framework for Twisted Bilayer Photonic Crystal with Interlayer Coupling and Far-Field Response}

\author{Shupeng Xu}
\author{Dun Wang}%
\author{Ritesh Agarwal}%
 \email{riteshag@seas.upenn.edu}
\affiliation{%
 Department of Materials Science and Engineering, University of Pennsylvania, Philadelphia, Pennsylvania 19104, USA
}%

%
%

\date{\today}

\begin{abstract}
We develop a general theory for twisted bilayer photonic crystals that takes into account both far-field response 
and near-field coupling.
The theory is based on the framework of a generalized Rayleigh-Schr\"odinger perturbation 
theory for non-Hermitian Hamiltonians.
A universal form for interlayer coupling is derived, which relates the hopping strength to
the Fourier transforms of the Wannier functions in the single layer photonic crystal.
For low energy states 
at the K point in hexagonal lattices, the interlayer coupling reduces to that in the Bistritzer-MacDonald model for graphene.
As an example, we study a twisted bilayer photonic crystal slab with air holes arranged in a honeycomb lattice in each layer. 
The first order solution of our model predicts a four-fold band splitting in the far-field spectrum 
compared to the single-layer case, which is confirmed by numerical simulations.
Moreover, our theory reveals that for low energy states at K points, scattering towards the $\Gamma$ point via the moir\'e potential
is suppressed.  
Based on our theory, we propose a wide-angle, high-Q tunable flat band cavity by combining
the bilayer at a large twist angle with a Brillouin-zone-folding perturbation within each layer.
The cavity behaves like a collection of quasi-bound states in the continuum
with a divergent density of states, 
with potential applications in nonlinear optics, lasing and quantum optics.
\end{abstract}

\maketitle


In recent years, twisted moir\'e systems have garnered intense interest across a diverse range
of research communities. 
When two layers with similar lattice structures are stacked with a twist angle, 
they form a moiré superlattice with enlarged periodicity. 
This large length scale modulation dramatically alters the band structure, 
enabling the emergence of flat bands. 
In twisted van der Waals materials, the kinetic energy of electrons is quenched, 
and many-body interactions dominate their physical properties. 
This leads to a plethora of exotic 
states such as correlated insulators, superconductors, topological insulators
and observation of novel phenomena such as fractional quantum anomalous Hall
and nonlinear Hall effects~\cite{cao2018correlated,cao2018unconventional,park2023observation,zeng2023thermodynamic,kang2024evidence,ji2024local,ji2024opto}.
In photonic systems, twisted bilayer photonic crystals~(TBPCs) enable localized optical modes and 
tunable resonances, 
leading to moir\'e lasers \cite{ouyang2024singular,luan2023reconfigurable,mao2021magic},
quasi-bound states in the continuum~(quasi-BICs)  \cite{huang2022moire}, sensors \cite{tang2025adaptive} and 
other novel devices \cite{lou2024free,lou2022tunable}.

Theoretical studies have been conducted to understand the properties of TBPCs,
which can be broadly classified into two categories.
The first category includes photonic analogs of the Bistritzer-MacDonald model \cite{bistritzer2011moire} originally developed for twisted bilayer graphene (TBG).
These models~\cite{dong2021flat} along with experiments~\cite{huang2022moire,mao2021magic},
can solve for band structures in the moir\'e Brillouin zone; 
however, they offer no insights into the far-field optical response.
Moreover, these models are difficult to generalize to photonic structures that lack the analogy of tightly bound 
atomic orbitals, such as dielectric slabs with air hole arrays,
where the hopping integral of orbitals is not defined.
The second category, exemplified by Ref.~\cite{lou2021theory}, focuses on solving the far-field response,
but neglects the near-field coupling between the two layers.
For moir\'e photonic crystals, both far-field response and near-field coupling are critical.
The former determines the optical functionalities relevant for many device applications, 
while the latter enables the flat band physics.
A theoretical framework that integrates both effects is therefore needed to fully understand these systems.

Here we present a unified theoretical framework that incorporates both far-field response and near-field coupling in TBPCs.
We start with a toy model to gain some intuition about the far-field response.
Consider an effective non-Hermitian Hamiltonian for a photonic crystal subjected to a perturbation with moir\'e periodicity:
\begin{equation}
  \mathcal{\hat{H}=\hat{H}}_0 + \mathcal{\hat{V}},
  \label{eq: toy}
\end{equation}
where $\mathcal{\hat{H}}_0$ has unperturbed solutions $\mathcal{\hat{H}}_0|\psi_{0,\mathbf{k}}\rangle=E_{0,\mathbf{k}}|\psi_{0,\mathbf{k}}\rangle$,
and $\langle\psi_{0,\mathbf{k}}|\mathcal{\hat{V}}|\psi_{0,\mathbf{p}}\rangle=V_\mathbf{kp}\delta_{\mathbf{k,p+G}_m}$ obeys momentum conservation,
in which $\mathbf{G}_m$ is the moir\'e reciprocal lattice vector, 
and $\mathbf{k}$ and $\mathbf{p}$ are momenta in K space. 
Since exceptional points are generally rare in dielectric photonic crystals at general momenta, 
this allows for the following perturbative solution in Rayleigh-Schrödinger series~\cite{chen2025non,buth2004non}
\begin{equation}
  |\psi_\mathbf{k}\rangle = |\psi_{0,\mathbf{k}}\rangle + \sum_{\mathbf{G}_m} \frac{V_{\mathbf{k,k+G}_m} |\psi_{0,\mathbf{k+G}_m} \rangle}{E_{0,\mathbf{k+G}_m}-E_{0,\mathbf{k}}} + \cdots .
  \label{eq: perturbation}
\end{equation}
A solution in this form enables us to keep track of the far-field response instead of folding every state into the moir\'e Brillouin zone.
In the far field, the perturbative corrections radiate at momenta $\mathbf{k+G}_m$,
which show up as resonant features in experiments.
The net result is the emergence of satellite peaks due to the reduced translational symmetry.  
We note that, by appropriately choosing the unperturbed Hamiltonian $\mathcal{H}_0$, 
this result is reminiscent of the corresponding relation derived in Ref.~\cite{lou2021theory}.
However, as we will show later, instead of the satellite peaks, 
the most prominent feature of TBPCs is a band splitting which arises from degenerate perturbation theory.

Having analyzed the toy model for the far-field response under moiré perturbation, we now turn to the actual TBPC system.
The system is governed by the following Hamiltonian acting on the magnetic field $\mathbf{H(r)}$:
\begin{equation}
  \mathcal{\hat{H}} =  \nabla \times \left( \frac{1}{\epsilon(\mathbf{r})} \nabla \times \right),
\end{equation}
where $\epsilon(\mathbf{r})$ is the permittivity of the system.
The Hamiltonian can be partitioned into three parts 
\begin{equation}
  \mathcal{\hat{H}}= \mathcal{\hat{H}}_0 + \mathcal{\hat{V}}_1 + \mathcal{\hat{V}}_2,
\end{equation}
such that $\mathcal{\hat{H}}_0$ accounts for the vacuum background, 
and $\mathcal{\hat{H}}_0 + \mathcal{\hat{V}}_i$ equals to the Hamiltonian of the $i$th layer.
This theory takes the form of a tight binding model for two 2D potentials separated in the z-direction, 
in which the dominant term to consider is the interlayer coupling (see Supplemental Material for details):
\begin{align}
  \begin{split}
    \langle \psi_1 | \mathcal{\hat{H}} | \psi_2 \rangle & \approx \langle \psi_1 | \mathcal{\hat{V}}_2 | \psi_2 \rangle \\
    & \propto \int \mathbf{E}_1^*(\mathbf{r}) \Delta \epsilon_2 (\mathbf{r}) \mathbf{E}_2 (\mathbf{r}) d\mathbf{r},
    \label{eq: inter coupling 1}
  \end{split}
\end{align}
where $|\psi_i\rangle$ and $\mathbf{E}_i (\mathbf{r})$ denote the eigenstate and the corresponding electric field distribution
for the $i$th layer, respectively;
$\Delta \epsilon_2 (\mathbf{r})$ is the change in permittivity induced by the second layer.
Eq.~(\ref{eq: inter coupling 1}) takes the same form as the hopping integral in coupled mode theory~\cite{yariv2003coupled}, 
where $\Delta \epsilon (\mathbf{r})$ can be interpreted as an optical potential.

Since the Hamiltonians for individual layers are periodic, their solutions are Bloch waves. 
This allows us to write the electric fields of the eigenstates in the first and second layer as 
\begin{equation}
  |\psi_\mathbf{k}\rangle \big|_\mathbf{E} =\frac{1}{\sqrt{N}} \mathbf{u} (\mathbf{r})e^{i\mathbf{kr}}
\end{equation}
and
\begin{equation}
  |\psi_\mathbf{p}\rangle \big|_\mathbf{E} =\frac{1}{\sqrt{N}} \mathbf{v} (\mathbf{r})e^{i\mathbf{pr}},
\end{equation}
respectively, where $\mathbf{u(r)}$ and $\mathbf{v(r)}$ are the periodic parts of the Bloch functions,
and $\mathbf{k}$ and $\mathbf{p}$ are the in-plane Bloch vectors for each layer, with zero out-of-plane components.
Plugging them into Eq.~(\ref{eq: inter coupling 1}) and performing Fourier transform in the x-y plane, we have:
\begin{align}
  \begin{split}
  \langle \psi_\mathbf{k} | \mathcal{\hat{V}}_2 | \psi_\mathbf{p} \rangle 
  & = \frac{1}{N} \int e^{i(\mathbf{p-k})\mathbf{r}} \mathbf{u} (\mathbf{r})^* \Delta\epsilon_2(\mathbf{r}) \mathbf{v} (\mathbf{r}) d\mathbf{r} \\
  & = \Omega \sum_{\mathbf{G,G'}} \delta_{\mathbf{k+G, p+G'}} \int \tilde{\mathbf{u}}^*(\mathbf{G}) \tilde{\mathbf{V}} (\mathbf{G'}) dz,
  \label{eq: inter coupling 2}
  \end{split}
\end{align}
where $\Omega$ is the unit cell area, $\mathbf{G}$ ($\mathbf{G'}$) is the reciprocal lattice vector of the first~(second) layer, 
$\tilde{\mathbf{u}}(\mathbf{G})$ is the Fourier transform of $\mathbf{u}(\mathbf{r})$,
and $\tilde{\mathbf{V}}(\mathbf{G'})$ is the Fourier transform of $\Delta\epsilon_2(\mathbf{r})\mathbf{v}(\mathbf{r})$.
Eq.~(\ref{eq: inter coupling 2}) is interpreted as follows:
a state with momentum $\mathbf{k}$ in the first layer can couple to a state with momentum $\mathbf{p=k+G-G'}$ in the 
second layer,
with the coupling strength determined by the product $\mathbf{u^*(G)V(G')}$.
This selection rule in momentum space is the same as that in typical continuum theories for moir\'e systems~\cite{bistritzer2011moire,dong2021flat}.

We now analyze the Fourier transform in Eq.~(\ref{eq: inter coupling 2}) to gain further insight into
the interlayer coupling.
We start with the Fourier transform of the periodic part $\mathbf{u(r)}$ of the Bloch function.
$\mathbf{u(r)}$ can be expressed in terms of Wannier function:
\begin{equation}
  \mathbf{u}_{\mathbf{k}_0}\mathbf{(r)} = e^{-i\mathbf{k}_0\mathbf{r}}|\psi_\mathbf{K}\rangle = 
  \frac{1}{\sqrt{N}}\sum_{\mathbf{R}_i} e^{i\mathbf{k}_0(\mathbf{R}_i-\mathbf{r})} \mathbf{w}(\mathbf{r-R}_i),
  \label{eq: wannier}
\end{equation}
where $\mathbf{R}_i$ denotes the discrete lattice vectors in real space,
$\mathbf{k}_0$ labels the Bloch vector of the state,
and $\mathbf{w}(\mathbf{r}-\mathbf{R}_i)$ is the Wannier function at lattice site $\mathbf{R}_i$.
The Fourier transform of Eq.~(\ref{eq: wannier}) is given by
\begin{equation}
  \tilde{\mathbf{u}}_{\mathbf{k}_0}(\mathbf{k})=\sqrt{N} \sum_{\mathbf{G}} \delta_{\mathbf{k,G}} \tilde{\mathbf{w}}(\mathbf{k}+\mathbf{k}_0),
  \label{eq: fourier}
\end{equation}
in which $\mathbf{\tilde{w}}(\mathbf{k})$ is the Fourier transform of the Wannier function.
The Wannier function can be constructed to follow the symmetry of the unit cell and 
have symmetric Fourier transform around $\mathbf{k=0}$.
Eq.~(\ref{eq: fourier}) shows that $\tilde{\mathbf{u}}_{\mathbf{k}_0}$
is nonzero only at reciprocal lattice vectors $\mathbf{G}$, 
and the value is determined by the envelope function $\tilde{\mathbf{w}}(\mathbf{k}+\mathbf{k}_0)$, 
which is the Fourier transform of the Wannier function shifted to $-\mathbf{k}_0$.
For a system whose single layer is arranged in a hexagonal lattice (as shown in Fig.~1(a) and (b)),
Fig.~1(c) illustrates Eq.~(\ref{eq: fourier}) for a state with $\mathbf{k}_0$ at the K point.
The above discussion assumes a single-band senario in Eq.~(\ref{eq: fourier}), 
but the generalization to the multi-band case is straightforward.
We note that the behavior of $\mathbf{\tilde{w}(k)}$ is directly related to the real-space characteristics of $\mathbf{\tilde{w}(r)}$;
in particular, more spatial oscillations in the Wannier function correspond to higher amplitudes at large momenta in $\mathbf{\tilde{w}(k)}$.
Additionally, such oscillatory states are typically found in higher energy bands due to the spatial derivatives in the Hamiltonian.
Thus, we expect $\mathbf{\tilde{u}(k)}$ to take large amplitudes at high-order $\mathbf{G}$ points in high-energy bands,
while in low-energy bands, $\mathbf{\tilde{u}(k)}$ is significant only at a few of the lowest-order $\mathbf{G}$ points. 
We now consider the second component $\mathbf{\tilde{V}(G')}$ in the interlayer coupling.
$\mathbf{\tilde{V}(G')}$ is given by the convolution of the Fourier transforms 
of its two components, $\Delta\epsilon_2(\mathbf{r})$ and $\mathbf{v(r)}$, 
in which the latter follows a similar form as Eq.~(\ref{eq: fourier}).
The Fourier transform of $\Delta\epsilon_2(\mathbf{r})$ is straightforward,
and it has significant values at least up to first order in $\mathbf{G'}$.
Therefore, we do not assume any particular pattern for $\mathbf{\tilde{V}(G')}$, 
which generally takes nonzero values at the relevant $\mathbf{G'}$.
Consequently, the allowed interlayer scattering is primarily controlled by $\mathbf{\tilde{u}(G)}$.

\begin{figure}[h]
  \hspace*{-0.8cm}
\includegraphics[width=1.15\linewidth]{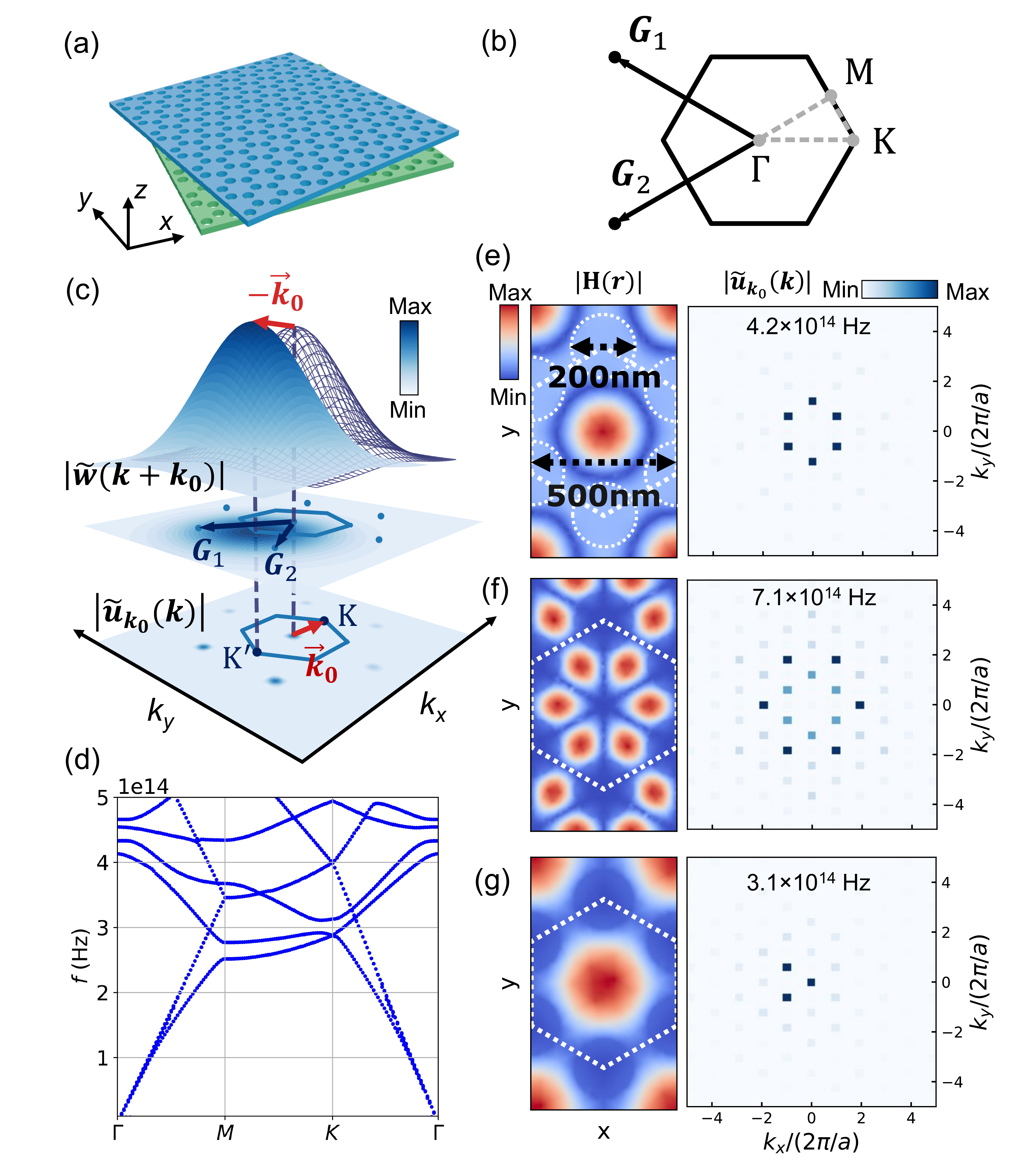}
\caption{\label{fig:epsart} 
(a) A schematic of a twisted bilayer photonic crystal slab with hexagonal unit cell in each layer. 
(b) A schematic of the Brillouin zone for the hexagonal lattice in a single layer. 
$\mathbf{G}_1$ and $\mathbf{G}_2$ are the reciprocal lattice vectors; $\Gamma$, $M$ and $K$ are high symmetry points in the K space. 
(c) A schematic for $\tilde{\mathbf{u}}_{\mathbf{k}_0}(\mathbf{k})$ with $\mathbf{k}_0$ at 
the K point of the Brillouin zone. 
The surface plot shows the envelop function $\tilde{\mathbf{w}}(\mathbf{k}+\mathbf{k}_0)$, 
which is obtained by shifting $\tilde{\mathbf{w}}(\mathbf{k})$ (the wireframe centered at $\mathbf{k=0}$) 
to the opposite K point. 
The middle plane shows the magnitude of $\tilde{\mathbf{w}}(\mathbf{k}+\mathbf{k}_0)$. 
The blue hexagon denote the First Brillouin zone, and the blue dots mark the reciprocal lattice vectors. 
The lower plane shows the magnitude of $\tilde{\mathbf{u}}_{\mathbf{k}_0}(\mathbf{k})$.
(d) Band structure for the single layer photonic crystal slab. 
(e-g) Real space field distribution (left panels) and norm of $\tilde{\mathbf{u}}_{\mathbf{k}_0}(\mathbf{k})$ 
(right panels) in K space for a low energy state at $\Gamma$ point ($f=4.232\times10^{14} \text{Hz}$),
a high energy state at $\Gamma$ point ($f=7.085\times10^{14} \text{Hz}$), 
and a low energy state at $K$ point ($f=3.142\times10^{14} \text{Hz}$), respectively. 
$\tilde{\mathbf{u}}_{\mathbf{k}_0}(\mathbf{k})$ is calculated at a plane 10 nm away from the slab. 
The left panel of (e) shows the geometry of the single layer slab. 
White dotted circle mark the air holes with diameter of 200 nm; rest of the area is silicon nitride. 
The lattice constant of the lattice is 500 nm.
}
\end{figure}

To quantify the interlayer coupling, we performed numerical calculations for a single-layer photonic crystal slab.
The system consists of a 200-nm-thick silicon nitride slab with an air-hole array arranged in a honeycomb lattice.
The air holes have diameter of 200 nm, and the lattice constant of the hexagonal lattice is 500 nm,
as shown in the left panel of Fig.~1(e).
All subsequent numerical calculations assume the same single-layer configuration, 
unless stated otherwise.
The band structure of the single layer is shown in Fig.~1(d).
We selected some representative states to calculate their field distributions and $\mathbf{\tilde{u}(k)}$ outside of the slab. 
In Fig.~1(e) and (f) we plot the results for a low-energy and a high-energy state at $\Gamma$ point, respectively.
As shown, $|\mathbf{\tilde{u}(k)}|$ takes nonzero values at discrete $\mathbf{G}$ points forming a triangular lattice.
For the low-energy state, $|\mathbf{\tilde{u}(k)}|$ has significant values only at the first-order lattice vectors, 
while for the high-energy state, it peaks at the second-order $\mathbf{G}$,
consistent with our previous discussion.
Notably, both states have $|\mathbf{\tilde{u}(k)}|=0$ at $\mathbf{G=0}$, 
which is related to the vortex center of the far-field polarization and optical BICs \cite{zhen2014topological}.
We also examined a low-energy state at K point, as shown in Fig.~1(g).
In this case, $|\mathbf{\tilde{u}(k)}|$ is significant only at three first order $\mathbf{G}$ points centered around the opposite K point,
which agrees with Fig.~1(c).
We note that these three leading order scatterings match exactly with the Bistritzer-MacDonal model,
so that our model reduces to the correct limit in the low energy regime.

To put our discussion into perspective, 
our results for the interlayer coupling generalize previous theoretical works on photonic systems of the first category (TBG analogies).
Our framework provides several advantages.
First, we do not rely on the existence of tightly bounded atomic orbitals to calculate the interlayer coupling strength.
Instead, our formalism works for all periodic systems 
including photonic crystal slabs with air holes.
Second, the procedure we developed to obtain $\mathbf{\tilde{u}(k)}$ is not limited to the K point in hexagonal lattices.
For example, results in Fig.~1(e) can be used to formulate theories for quadratic dispersions at $\Gamma$ point. 
Lastly, we obtained new insights into the nature of higher-order scattering.
In Ref.~\cite{dong2021flat}, it is argued that as interlayer separation can be small in photonic systems, 
higher order scattering needs to be included.
In our framework, this effect corresponds to the $z$ integral in Eq.~(\ref{eq: inter coupling 2}):
since higher-order component corresponds to evanescent wave with smaller decay length, 
it can only contribute at small interlayer spacing.
However, we showed that, besides interlayer spacing, 
the magnitude of higher-order scattering strongly depends on the energy of the state.
In the low-energy regime,
higher-order components can be absent in the Wannier function to start with,
while at high-energy regime, high-order scattering can be the dominant contribution, as shown in Fig.~1(e) and (f).
With the results derived above, it is straightforward to formulate a TBG-like theory and solve for the photonic band structure.
We note that so far, we only considered the magnitude of the interlayer couplings.
The phases of the couplings may lead to interesting consequences
that are unique for photonic systems,
which we leave for future research.

In the following, we examine the far-field properties of TBPCs.
First, we notice that for a single layer, the resonant frequencies at a certain wave vector $\mathbf{k}$
are in the vicinity of 
\begin{equation}
  \omega = \frac{c}{\sqrt{\epsilon_\text{eff}}} \left| \mathbf{k+G} \right|,
  \label{eq: effective}
\end{equation}
where $\epsilon_\text{eff}$ is the effective permittivity of the single layer photonic crystal.
In Fig.~2(a) we show the schematic of an iso-frequency slice of the resonant states for a decoupled twisted bilayer.
The dotted circles indicate the resonant states, 
and different colors correspond to different layers.
The circles originate from the linear dispersion in Eq.~(\ref{eq: effective})
and are centered around the twelve first-order $\mathbf{G}$ points of the two layers.
These circles account for the first set of resonances above the light cone.
Now we consider the coupling between the two layers.
According to Eq.~(\ref{eq: inter coupling 2}),
modes in different layers are coupled when their momenta satisfy $\mathbf{p=k+G-G'}$.
We define $\mathbf{G}_m=\mathbf{G-G'}$ as the moir\'e $\mathbf{G}$ vector 
(which also applies for incommensurate angles),
and notice that for every pair of $\mathbf{G}$ and $\mathbf{G'}$ related by a twist rotation, 
the states on the corresponding circles are separated exactly by $\mathbf{G}_m$,
as shown in Fig.~2(a).
Thus, for every state on a pair of circles, we have a degenerate perturbation problem with basis states $|\psi_\mathbf{k}\rangle$
and $|\psi_{\mathbf{p=k+G}_m}\rangle$.
The two zeroth-order corrected eigenstates have the same weight on both layers 
and take the form $|\psi_\mathbf{k}\rangle + e^{i\phi} |\psi_\mathbf{p}\rangle$,
with an energy splitting between each other.
Both eigenstates radiate equally into the circles corresponding to the two layers.
As a result, every circle will be split into a doublet in the far-field spectrum, which we show in Fig.~2(b).

\begin{figure}[h]
  \hspace*{-0.75cm}
\includegraphics[width=1.05\linewidth]{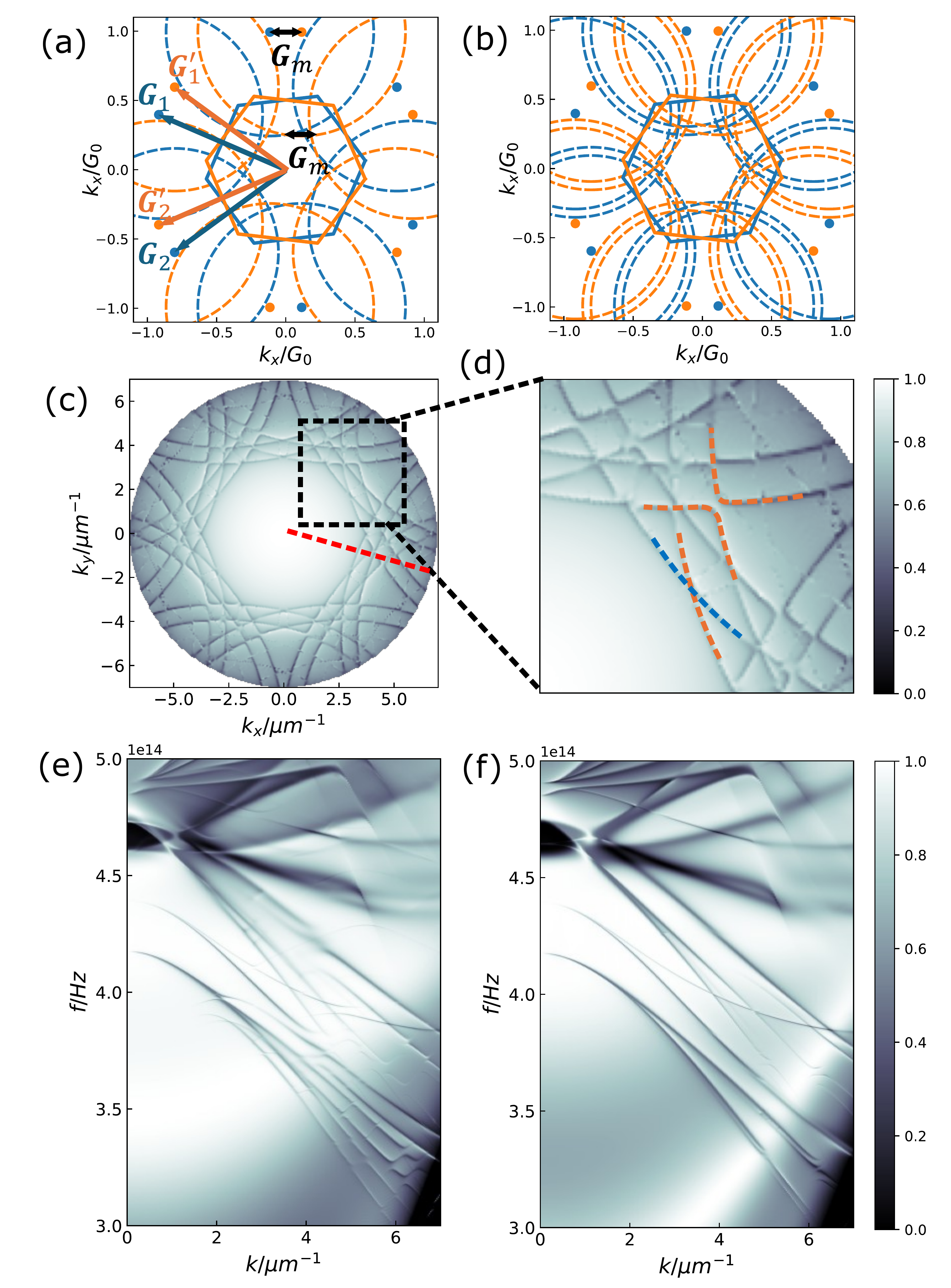}
\caption{\label{fig:2} (a) A schematic of an iso-frequency slice of the far-field spectrum for a decoupled TBPC. 
The solid hexagons represent the first Brillouin zone, the dots are the first-order reciprocal lattices, 
and the dotted circles denote the resonant states. 
Different colors correspond to different layers. 
(b)~A schematic of the far-field spectrum of a coupled TBPC. 
The circles split into doublets due to interlayer coupling. 
(c)~Numerical result of the transmission spectrum for a coupled TBPC. 
(d)~A zoomed-in plots on the crossing regions from (c). 
The color of the dotted lines follow from (b). 
(e)~and~(f)~Transmission spectrum at 15 degrees from the $x$-axis (red dotted line in (c)). 
Interlayer spacings: 300 nm (e) and 500 nm (f).}
\end{figure}

As an example, we numerically calculated the transmission spectrum of a TBPC with 21.8 degrees twist angle 
(chosen for computational convenience) and 200 nm interlayer spacing. 
From the iso-frequency slice (Fig.~2(c)),
it can be seen that the resonances are grouped in multiple circular doublets with considerably large splitting, 
which matches well with Fig.~2(b).
Besides the splitting, other fine features can be observed.
In Fig.~2(d) we show a zoomed-in plot of the circle crossings.
We observe that anti-crossings occur only at the intersections of circles with the same color
and are absent at crossings between circles of different colors.
This implies that the zeroth-order interlayer coupling ($\mathbf{G}_m=\mathbf{0}$) is small.
Such behavior is reminiscent of Fig.~1(e), 
where first-order scattering dominates and zeroth-order contributions are suppressed.
The absence of the anti-crossings is intimately related to optical BICs, which have a topological origin.
For all states not too far away from $\Gamma$ point 
on a band with symmetry protected BIC,
the zeroth-order scattering will remain suppressed 
since BIC forces the envelope function $\mathbf{\tilde{w}}(\mathbf{k})$ to be zero at $\mathbf{k=0}$.
We note that similar effect was argued in Ref.~\cite{huang2022moire},
where the suppressed zeroth-order scattering was attributed to the large interlayer spacing and the low radiation loss of BICs.
Here, we provided a direct numerical observation of this effect and 
generalized it to small interlayer spacing in which evanescent coupling dominates.
This interplay between interlayer coupling and BICs, as captured by our theory, 
is a unique feature of photonic systems, 
where the underlying physics is governed by Maxwell's equations.
Lastly, in Fig.~2(e) and (f), we plot the transmission spectrum of the system 
along an axis 15 degrees from the x-axis (dotted line in Fig.~2(c)).
Compared to Fig.~1(d), a single band in the monolayer splits into four bands in a twisted bilayer system
at the far field,
even though the number of resonant states only doubled. 
It can be seen from Fig.~2(c) that this four-band splitting is a direct consequence of the splitting of the circles,
as the dotted line produces four intersections at the innermost hexagon emerging from the circles.
Moreover, the splitting increases as the interlayer spacing becomes smaller, 
which is consistent with the interpretation that it arises from evanescent coupling of the fields.

Besides solving the far-field spectrum, 
our theory has practical implications for potential device applications.
For example, 
previous studies have reported that small-angle moiré photonic structures can host 
localized optical modes with extremely large out-of-plane quality factors~\cite{ouyang2024singular,luan2023reconfigurable,mao2021magic}. 
The physical origin of this effect is that, for small twist angles, the moiré reciprocal vectors $\mathbf{G}_m$ are small. 
Consequently, Dirac cones located below the light line can couple to the far field only through 
high-order perturbative processes in Eq.~(\ref{eq: perturbation}), which strongly suppresses radiative loss. 
Importantly, in these works the moiré potential was effectively implemented by 
overlapping the motifs of two lattices and projecting it to a single layer.
Therefore, the Hamiltonian more closely resembles Eq.~(\ref{eq: toy}),
where the moir\'e potential scatters in all directions and may lead to radiative loss at large twist angles.
In contrast, in our theory, the moiré physics arises from interlayer coupling that follows Eq.~(\ref{eq: inter coupling 2}),
with only three leading-order terms for low energy states at K point.
In Fig.~3(a) we show a schematic of these scatterings
for a TBPC system.
The solid arrows represent the two moir\'e scatterings with nonzero momenta, 
which are given by $\mathbf{G}_m = \mathbf{G-G'}$;
and the black circle shows the light line at the frequency of single-layer Dirac cone.
As can be seen, none of the leading-order scatterings can bring the Dirac cone above the light line
even at large twist angle.
Thus, we expect the high quality factors can be accessed even at large twist angles, 
which dramatically shrinks the size of the moir\'e unit cell.

\begin{figure}[h]
  \hspace*{-0.7cm}
\includegraphics[width=1.1\linewidth]{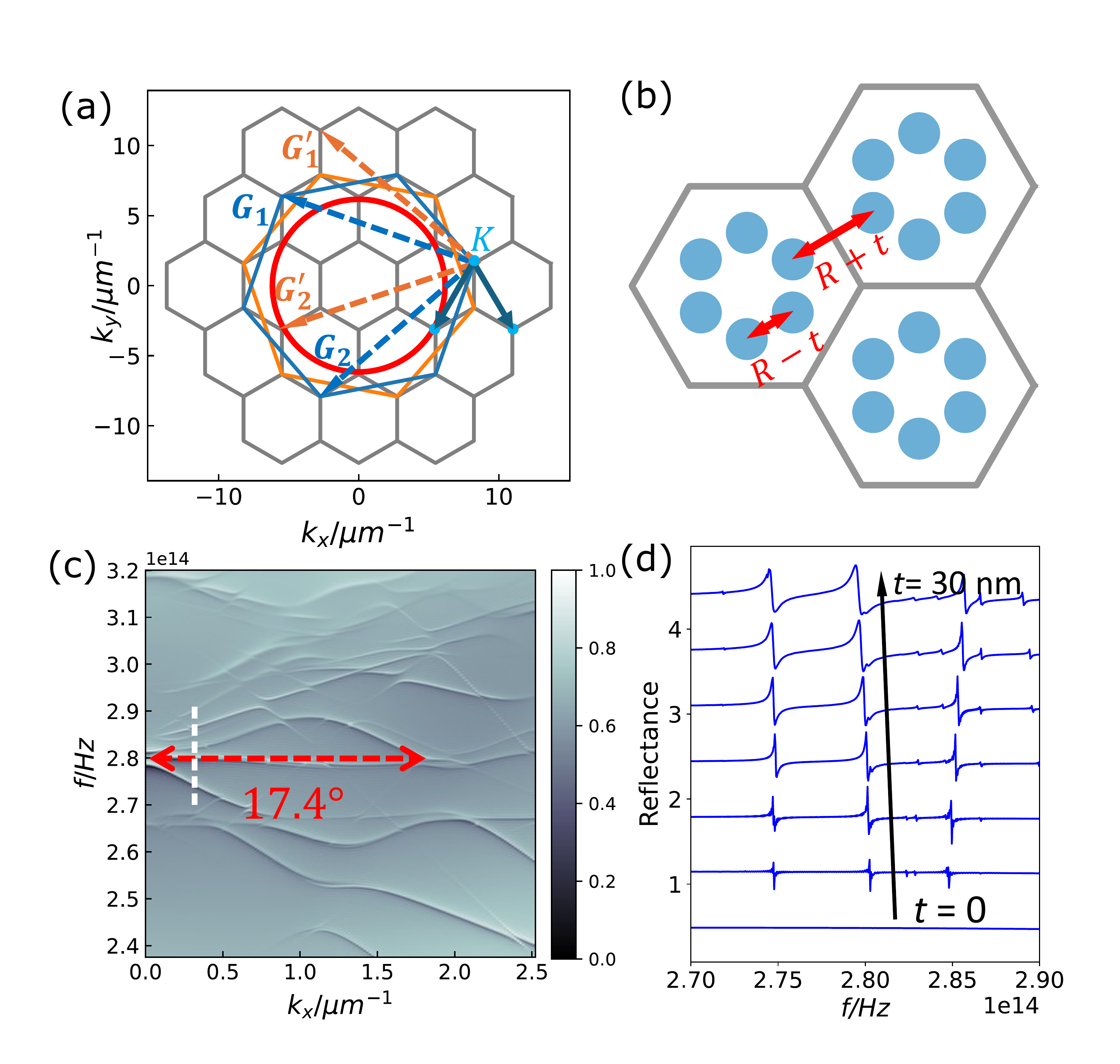}
\caption{\label{fig:3} (a) A schematic of the leading-order scattering for low energy state at K point for a TBPC. 
Solid (blue and orange) large hexagons are the single-layer first Brillouin zones 
and gray small hexagons are moire Brillouin zones; 
the twist angle is $21.8^\circ$. 
Dotted arrows are the reciprocal lattice vectors of the single layers; 
the solid arrows shows the leading-order scattering at K point. 
The red circle is the light line at the energy of single-layer Dirac cone. 
(b) A schematic of the Wu-Hu model in real space (Brillouin-zone-folding perturbation), 
the air holes are displaced by $t$. 
The hexagons are the enlarged unit cells. $R=a/\sqrt{3}$ is the distance between the nearest air holes of 
the unperturbed honeycomb lattice, whose lattice constant is $a$. 
(c) Transmission spectrum for the TBPC with 130 nm interlayer spacing, $21.8^\circ$ twist angle and $t$=30 nm. 
A flat band that covers a wide range of incidence angle is marked by the red dotted arrow. 
(d) Transmission spectra at the line cut shown as white dotted line in (c) as a function of the perturbation strength t. 
Resonances get sharper as $t$ decreases while they disappear at $t$ = 0.}
\end{figure}

To demonstrate this idea, we propose a tunable, high-Q flat band cavity based on large-angle TBPCs.
We combine the moir\'e structure with a Brillouin-zone-folding perturbation in each layer
in order to tune the Q factors of the flat band.
The single-layer perturbation is added in the same way as in the 
Wu-Hu model of photonic crystals~\cite{wu2015scheme,xu2023absence,liu2020generation}.
Specifically, the unit cell of the honeycomb lattice of the single layer is expanded to include six atomic sites, 
and their positions are tuned away or towards the center of the unit cell. 
We denote the distance by which the air holes are moved as $t$, and a schematic of the perturbation is shown in Fig.~3(b).
In Fig.~3(c) we show the transmission spectrum of the system with 130 nm interlayer spacing, 21.8 degrees twist angle and $t$ of 30 nm.
A flat band that covers a wide range of incident angles can be observed. 
In addition, the Q factors of the flat band modes solely depend on the magnitude of the perturbation $t$.
In Fig.~3(d) we plot the transmission spectrum of a particular flat band mode with different $t$.
As shown, the resonant peak gets sharper as $t$ approaches zero; 
and the far-field feature completely vanishes at $t=0$, which indicates an infinite Q factor.
The cavity we proposed thus behaves like a collection of quasi-BICs that resembles the Brillouin-zone-folding BICs~\cite{wang2023brillouin},
but with divergent density of states due to the moir\'e structure.
The combination of tunable high Q factors and divergent density of states can lead to 
high nonlinearity enhancement and high Purcell factors, 
which have potential applications in nonlinear optics, quantum optics and lasing, etc.
Our proposal can be readily implemented in silicon photonics, 
a well-established experimental platform where significant progress in moiré photonics has already been demonstrated~\cite{tang2025adaptive,tang2023experimental}.

In conclusion, we have developed a unified theoretical framework to understand both the near-field coupling
and far-field properties of TBPCs.
We obtained a general expression of interlayer coupling,
from which the scattering strength can be obtained 
based on the single-layer eigenstate field distribution.
We further solved for the far-field radiation to leading order 
and predicted a four-band splitting in the transmission spectrum.
Our theory unifies and generalizes a variety of existing observations in the literature,
while providing new predictions with potential relevance for device applications.
The framework we established sets the stage for a broad range of future studies in moiré photonic systems, 
offering a versatile platform for exploring light-matter interactions, 
engineered band structures, novel topological effects, and beyond.

\textit{Acknowledgments}---This work was supported by the Office of Naval Research under Grant No. N00014-22-1-2378 
and by the National Science Foundation through the DMREF program (Grant No. NSF-2323468).
The authors are grateful to Eugene J. Mele for helpful discussions.
The authors also thank Flexcompute Inc. for providing access to 
the Tidy3D software used for all FDTD simulations in this work.

\textit{Data availability}---The simulation project and associated code are available in Ref.~\cite{tidy3d}. 
A Tidy3D software license can be obtained from Flexcompute Inc. to reproduce the simulation results.

\bibliography{apssamp}

\end{document}